# Sulfur Volcanism on Io

Kandis Lea Jessup, Dept. Space Studies, Southwest Research Institute, 1050 Walnut St., Suite 400, Boulder, CO 80302

John Spencer, Dept. Space Studies, Southwest Research Institute, 1050 Walnut St., Suite 400, Boulder, CO 80302

Roger Yelle, Lunar and Planetary Laboratory, University of Arizona, Tucson, AZ 85271

Pages: 44
Tables: 3
Figures: 7

Sulfur Volcanism on Io

Kandis Lea Jessup

Southwest Research Institute

1050 Walnut, Suite 400

Boulder, CO 80302

**Abstract:**

In February 2003, March 2003 and January 2004 Pele plume transmission spectra were obtained during Jupiter transit with Hubble's Space Telescope Imaging Spectrograph (STIS), using the 0.1" long slit and the G230LB grating. The STIS spectra covered the 2100-3100 Å wavelength region and extended spatially along Io's limb both northward of Pele. The $S_2$ and $SO_2$ absorption signatures evident in the these data indicate that the gas signature at Pele was temporally variable, and that an $S_2$ absorption signature was present ~ 12° from the Pele vent near 6±5 S and 264±15 W, suggesting the presence of another $S_2$ bearing plume on Io. Contemporaneous with the spectral data, UV and visible-wavelength images of the plume were obtained in reflected sunlight with the Advanced Camera for Surveys (ACS) prior to Jupiter transit. The dust scattering recorded in these data provide an additional qualitative measure of plume activity on Io, indicating that the degree of dust scattering over Pele varied as a function of the date of observation, and that there were several other dust bearing plumes active just prior to Jupiter transit. We present constraints on the composition and variability of the gas abundances of the Pele plume as well as the plumes detected by ACS and recorded within the STIS data, as a function of time. We discuss the implications of these results for thermochemical conditions at the plume vents, and our understanding of plume eruption styles active on Io.

## 1.0 Introduction:

The Pele volcano produces one of Io's largest and most dynamic plumes. It is well known that sulfur dioxide ($SO_2$) gas is the dominant atmospheric species on Io, and that this gas ultimately originates from Io's active volcanic plumes. If uncontaminated, the $SO_2$ gas released in Io's volcanic eruptions will produce white surface deposits. However, red surface deposits are prominent in the ring encompassing the Pele vent (located at 18 S, 258 W) at a radius of 500-700 km, and in several narrow diagonal streaks extending directly from the Pele vent (Figure 1). Pele's brillant red surface coloration is consistent with short chain sulfur products (Nelson and Hapke 1978, Spencer *et al*. 1997a). Because these products are produced by the quenching of sulfur gas ($S_2$), sulfur gas was long suspected to be a major constituent of the Pele plume (McEwen and Soderblom 1983). Spectral observations obtained in 1999 with Hubble's Space Telescope Imaging Spectrograph (STIS) were the first to simultaneously identify and quantify the density of sulfur dioxide ($SO_2$) and $S_2$ gas (Spencer *et al.* 2000) within the Pele plume. From these observations an $S_2/SO_2$ ratio in the range of 8-30% was observed, confirming the dominance of $SO_2$ gas within the Pele plume, and that $S_2$ gas was a significant constituent of the Pele eruption.

INSERT

FIGURE 1

Galileo observations obtained during the same time period provided measurements of Pele's thermal signature revealing magma temperatures of 1600 K or higher. In particular, single temperature modeling of the Solid State Imager (SSI) (McEwen *et al.* 1998a) and the Near Infrared Mapping Spectrometer (NIMS) data (Lopes *et al*. 2001) initially provided magma temperatures of 1275±15 K and 1760±210 K, and an estimate of 1250-1450 K from the combined analysis of the SSI-NIMS data. Two-temperature models of the combined NIMS-SSI data returned estimates of 1077±13 K and 1390±25 K, with a need for >1600 K magma temperatures to fully explain the data (Davies et al. 2001). However, the most recent analysis of the SSI data (Keszthelyi *et al*. 2005) limits Pele's maximum magma temperature to 1600 K.

At these high temperatures the erupted gases reach thermodynamic equilibrium with each other and the adjacent magma; thus, measurement of the plume's gas composition can be used to probe magma chemistry and vent conditions. Exploiting this fact Zolotov and Fegley (2000) produced thermochemical models that used measurements of the $SO_2$/S/SO gas ratios derived from the 1996 HST / Faint Object Spectrograph (HST/FOS) observations of Pele (McGrath et al. 2000) to predict vent temperatures, and derive the oxygen fugacity of the magma at Pele. Combining these results with the $S_2/SO_2$ ratios derived from the 1999 HST/STIS observations, conditions at the vent were constrained to correspond to pressures of $3.8x10^{-6}$-$1.8x10^{-5}$ bar, temperatures of 1440±150 K and $SO_2/S_2$/SO/S mole fractions of 0.71-0.86/0.24-0.07/0.05-0.07/0.0022-.0027, additionally the composition of other minor gas species erupted at the vent were constrained (Zolotov and Fegley 2000-2001). Thus, the $S_2/SO_2$ ratio derived from the 1999 observations provided a critical component necessary for the constraint of the bulk chemistry and eruption conditions of the Pele plume.

However, Pele has long been known to be episodic and variable in her eruptive nature. Consequently, in February 2003, March 2003 and January 2004 additional Pele plume transmission spectra were obtained during Jupiter transit with STIS, using the 0.1" long slit and the G230LB grating. The observations were obtained on multiple dates in order to determine if the plume composition observed in 1999 was typical for the Pele plume, or to what degree the plume composition at Pele was temporally variable (see Table I). Additionally, we planned to obtain longer exposure (higher signal to noise) observations of the Pele plume so that the compositional constraints would be improved. Contemporaneous with the spectral data, UV and visible-wavelength images of the plume were obtained in reflected sunlight with the Advanced Camera for Surveys (ACS) prior to Jupiter transit (Table II). From these observations plumes erupting along Io's limb would be evident due to the backscattering of the solar light by dust in the plumes. By obtaining imaging observations contemporaneously with the spectral data the relative abundance of gas and dust within the Pele plume could be documented. Because Galileo had difficulty observing the Pele plume at visible wavelengths (McEwen et al. 1998b), it has been speculated that the Pele plume is somewhat stealthy in nature (i.e. it is low in dust content and high in gas content) (Johnson *et al.* 1995). In spite of this, on each of the dates of the HST observations $S_2$ gas and dust scattering were detected near the Pele vent (Fig. 2). Additionally, dust scattering was revealed during the 2003 observations at several locations not coincident with the Pele vent indicating that during those observations several other dust bearing plumes were active on Io just prior to Jupiter transit (Fig. 2).

INSERT FIG. 2

INSERT

TABLES I & II

We present a detailed analysis of the $SO_2$ and $S_2$ gas densities present at each of the locations where dusty plumes were coincidentally detected, and discuss the implications of these results relative to the observed dusty plume activity. We also discuss the implications of the gas density ratios observed at Pele in 2003 and 2004, relative to previous plume observations and the magma conditions first derived by Zolotov and Fegley (2000). Lastly, we discuss the significance of the detection of $S_2$ gas in regions other than Pele, and the implication of these detections on our understanding the style of eruption associated with each volcano.

## 2.0 Observation Description and Data Reduction Methods:

### *2.1STIS imaging:*

HST/STIS spectra of the Pele plume were obtained on 6 separate dates between October 1999 and January 2004. In each case, the spectra were obtained by using the 0.1" slit in conjunction with the G230LB grating and the CCD detector, producing spectrally dispersed HST/STIS images at and above Io's limb at 260 W during Jupiter transit. Because the images were spectrally dispersed over the 2000-3100 Å wavelength region we were able to isolate the gas absorption signatures in Io's plumes via the transmission of Jovian light through the plumes. Using HST/STIS in this configuration we obtained an un-smoothed spectral resolution of 5.5 Å and a spatial resolution of 160 x 320 $km^2$/pixel on Io.

STIS exposures obtained in October 1999 were published in Spencer *et al*. (2000). On this date exposures were obtained at 5 separate slit positions, separated by the width of the slit, where an integration time of 255 s was utilized at each slit position, and the expected Pele vent location corresponded to the third (or central) slit position (Fig. 3), producing an image cube with two spatial dimensions and one spectral dimension. This paper will focus on the HST observing program completed in 2003 and 2004, wherein spectral observations of Pele were obtained on 5 separate visits. On these visits, two exposures were taken at each of 3 slit positions, separated by the width of the slit and the total integration time of the central slit position was increased to 990-1070 s (see Table I). The goal of this observing scheme was to increase the integration time at the Pele plume, while ensuring that in the event of a pointing error the integration time utilized in the adjacent slit positions would produce S/N levels sufficient for the detection of the $SO_2$ and $S_2$ gas signatures. In spite of our improved targeting plan, Fig. 4 shows that in Visits 1, 3, and 4, the Pele vent and absorption signature of the gases emitted from the vent, were located in the top row of the image cube where the total integration time was only 260 s (see Table I). Additionally, on Visit 2 because of targeting errors a positive detection of Io transiting Jupiter was recorded in only one of the slit positions. In this case the slit was positioned just above the vent of Pele, barely grazing Io; in the other two slit positions (not shown) only Jupiter was exposed. On Visit 5 the Pele plume was successfully centered in the slit position for which the long integration exposure was taken. For this visit we only show two of the slit positions because in the third position only Jupiter was detected.

INSERT

FIG. 3 & 4

At wavelengths below 3200 Å Io's albedo is significantly lower than that of Jupiter, and so Io appears dark against the Jovian disk and the location of the limb is readily apparent. Exploiting this fact, the location of Io's disk, and the corresponding latitude and longitude of each spatial pixel in our image cube was determined using a least squares fit of the spatial profile produced by the image of Io within the slit. Specifically, a blank model disk of Io's known radius was convolved with a two-dimensional Gaussian function and then fitted to the observed spatial profile of Io produced by absorption of Jovian light by Io's disk, using a least squares fit where disk location, brightness and the Gaussian width were free parameters The FWHM of the STIS/CCD PSF is nominally 2 pixels (or 0.1"), the actually gaussian fit to the observations ranged in FWHM from 1.7 to 2.2 pixels (or 0.17 to 0.22") depending on the date of the observation. Based on these fits Io's disk is blocked out and represented as pure white in Figures 3 and 4. We include in the image cube only those slit positions that show the attenuation of Jovian light due to Jupiter being transited by Io or the Pele plume, because only those images can be used to constrain the fit of the model Io disk to the observed Io disk. In Fig. 4 the location of the gas absorptions within the image cubes is shown by summing up in each spatial pixel the flux at the bottom of the strong $SO_2$ and $S_2$ absorption bands, and then dividing the resulting image by the image produced by summing of the flux in the continuum regions between the bands. This scheme worked well for the $S_2$ absorption bands, but did not unambiguously identify the occurrence of the $SO_2$ gas absorption due to the lower signal to noise ratio of the data at wavelengths less than 2400 Å.

Basic reduction of the image cubes consisted of the removal of blemishes produced by hot pixels, cosmic ray hits, and bad pixels. Two exposures were taken at each of the slit positions per date of observation. We identify and remove bad pixels in the image cube based on a pixel by pixel comparison of the paired exposures; i.e., bad pixels (those whose values were more than 50% higher or lower than that of other nearby pixels) identified in one exposure were replaced with the pixel value in the undamaged exposure pair. In those cases where a hot/bad pixel region was evident in both exposures, the affected pixels were replaced by an interpolation of the flux levels from the surrounding pixels.

Additionally, scattered light was removed from each pixel. The CCD detectors are more sensitive at longer wavelengths than in the ultraviolet. As a result of grating scatter, STIS/CCD observations of solar system objects that have significantly higher flux at longer wavelengths, tend to have high levels of scattered light at wavelengths shorter than 2100 Å due to the long-wavelength sensitivity of the CCD detectors. Scattered light levels within each spatial pixel were determined based on the relative difference between the depth of the Fraunhofer lines evident in the observed and solar (Woods *et al.* 1996) spectra throughout the region extending from 2100 to 3100 Å.

The transmission signature of the Pele plume, and other active volcanic regions were derived based on the ratio of a spectrum taken along Io's limb to that of Jupiter away from Io's limb in the spatial direction. Specifically, on Visits 1 thru 4 the absorption signatures generated by an erupting plume were evaluated based on the co-addition of the first 3 or 4 pixels that extended spatially beyond the modeled location of Io's limb (see Fig. 4). On Visit 5 when the plume spectrum was not to the side, but above the Io limb, the gas absorption signature was determined from the 4 darkest pixels located above the Io limb. As discussed above, the FWHM of the STIS/CCD PSF is nominally 2 pixels (or 0.1"). Because of HST's PSF Io's spectrum was also recorded in the near Io-limb pixels. The additional flux from Io within these pixels was estimated via a comparison of the known Jovian to solar ratio with the ratio of the flux observed near the Io-limb to the solar flux as function of wavelength over multiple broad (~ 250 Å) spectral intervals. The jovian spectrum was determined from the average of the first 10 pixels located a minimum of 3 pixels beyond the region where the plume absorption was detected or expected. Using this scheme gas detections were made along Io's limb at ~266 W longitude within pixels centered near 18 S (Pele), 6 S, 10 N, and 18 N latitude (Fig. 4 & 5).

***2.2ACS imaging:***

On each date of the STIS observations, Io was imaged in reflected sunlight in the UV and the visible using Hubble's Advanced Camera for Surveys approximately two hours before Jupiter transit at a sub-earth longitudes in the range of 155-166 W, so that the limb longitude corresponded to a range of 245-256 W. Observations were made at multiple wavelengths throughout the 2000-8000 Å region. Our discussion of the dust detections focuses primarily on observations made with the ultraviolet F220W (2050-2450 Å) and F250W (2450-3020 Å) filters, (see Table II) where dust in the plumes is the most conspicuous in backscattered sunlight due to the low UV albedo of Io's surface (c.f.

Spencer et al. 1997b, Spencer et al. 2000). The ultraviolet image exposures were taken in pairs so that hot pixels and cosmic rays could be easily removed. The removal of cosmic rays was a concern for those images taken with the F220W and F250W filters where exposure times significantly greater than 18 s were required for adequate signal-to-noise (S/N). In order to avoid the inadvertent removal of plume signatures the cosmic ray removal was not done through the HST pipeline, instead we systematically replaced those pixels whose flux levels were more than twice the value of the average of the surrounding pixels with the average value of the surrounding pixels.

The F250W filter bandpass extends from 2000 to 3000 Å, allowing a significant throughput of near-UV and visible light beyond 2500 Å. Similarly, albeit not by design, the F220W filter allows significant transmission of near-UV light longward of 2500 Å. Consequently, Io's prominent near-UV albedo patterns are apparent in the images. We found that by sharpening the contrast within the F220W and F250W images and then applying a hard stretch to these images the visibility of the plumes along Io's limb was enhanced. As Figure 2 shows dust plumes are evident at several locations along Io's limb. A more detailed reduction of the images including estimating and removing the contribution beyond 2500 Å in the F220W and F250W filters, and deconvolution of the images will be done in a later work.

**3.0 Spectral Model Methodology:**

Each of the transmission spectra are modeled solely as a function of the amount of $SO_2$ and $S_2$ gas and the temperature of the $SO_2$ gas. With the advent of high resolution $SO_2$ cross-section data at multiple temperatures (Rufus *et al.* 2003, Smith *et al.* 2002, Stark *et* *al*. 1999), we were able to simultaneously fit the $SO_2$ gas absorption signature as a function of temperature and the $S_2$ gas density assuming a single $S_2$ gas temperature of 250 K throughout the 2000 to 2950 Å wavelength region. The best $SO_2$ and $S_2$ gas densities and $SO_2$ gas temperature were determined by grid search using a least squares fitting algorithm, and assuming $SO_2$ gas temperatures in the range of 100-1000 K, $SO_2$ gas densities in the range of $10^{14}$ to $5x10^{17}cm^{-2}$, and $S_2$ gas densities in the range of $10^{14}$ to $10^{17}$ $cm^{-2}$. The 250 K $S_2$ absorption cross-section was chosen because of the availability of new high resolution data; however, high resolution results for the $S_2$ molecule at multiple temperatures were not available at the time of modeling. Derivation of the $SO_2$ absorption cross-section temperature dependence is provided in previous publications (c.f. Jessup *et al*. 2004).

INSERT

FIG. 5

**4.0 Observational Trends and Spectral Fitting Results:**

***4.1 Trends at Pele***

*4.1.1 Gas signatures near 260 W*

At Pele the $SO_2$ gas density varied from 1.5 to $4.0x10^{16}$ $cm^{-2}$ and the $S_2$ gas density varied from 1.0 to $3.7x10^{15}$ $cm^{-2}$, depending on the date of observation. Due to low S/N at short wavelengths, and the shallowness of the weak $SO_2$ band absorption at long wavelengths the $SO_2$ gas temperatures are not well constrained. For the most part, the observations are statistically well fit at any $SO_2$ gas temperature between 100 and 600K. The only exceptions being the 6 S spectrum obtained on Visit 3 and the 18 N spectrum obtained on Visit 4. In these cases, because the depth of the weakly absorbing long wavelength $SO_2$ bands was low and the slope of the transmission at short wavelengths was relatively high these spectra were only well fit at $SO_2$ gas temperatures of 300 K or less. Thus, for these two spectra we list in Table III and plot in Fig. 6 the results obtained assuming an $SO_2$ gas temperature range of 100-300 K. Although, the $SO_2$ gas temperature inferred from previous spectra of the Pele plume ranges from 280±50 K (McGrath et al. 2000) to 300-500 K (Spencer et al. 2000), the available $S_2$ cross-section data is valid for gas temperatures of 250 ± 25K. Therefore, for the remaining spectra, we list in Table III and plot in Fig. 6 the results obtained assuming an $SO_2$ gas temperature range of 200-300 K. We additionally note that for these cases, the $SO_2$ column density fit to the spectra increases at temperatures between 300 and 600 K, but remains within a factor of two of the results listed in Table III. Similarly, for this temperature range the uncertainty in the fit $S_2$ column densities increases by 30% of the values listed in Table III. These increases are due to the decrease in the $SO_2$ absorption cross-section band contrast at warmer temperatures, requiring a greater amount of gas to fill in the spectral signature observed at both short and long wavelengths.

The absolute minimum reduced chi-square value is calculated based on minimizing the difference between the observed and model spectra given the error within the data defined from Poisson statistics and the number of degrees of freedom in the fits. The number of degrees of freedom in the fits is definedas a function of the wavelength range over which the data are fit, and the number of free parameters. The uncertainty in the fitting results corresponds to the amount by which the fitted parameters can change and still produce a model spectrum whose deviation from the observed spectrum results in a reduced chi-square value that is within 1 sigma of the absolute minimum reduced chi-square value.

INSERT

FIG. 6 &

TABLE III

As Fig. 2 and 6 show, variable levels of gas absorption were observed near Pele on each of the 5 visit dates. In the 3.5 weeks between Visit 1 and Visit 3, the $SO_2$ gas density increased by a factor of 2 from a range of 0.7-2.7x$10^{16}$ $cm^{-2}$ to 3.5-5.3x$10^{16}$ $cm^{-2}$, however 1 week subsequent to that on Visit 4 the $SO_2$ density level returned to that observed on Visit 1 and was also observed at that level ~ 9 months later on Visit 5. $S_2$ gas was also positively detected above the Pele vent on all visits. The apparent $SO_2$ gas variability could either result from actual changes in the gas density or variations in our spatial sampling of the gaseous plume. Although we cannot definitely state whether pointing affected our ability to observe the region of strongest $SO_2$ gas absorption above the Pele vent, the observed variability was not directly correlated with the relative position of Pele in the slit. In spite of the observed variation in the $SO_2$ column density, the $S_2$ gas density levels remained fairly constant at 0.7-2.0x$10^{15}$ $cm^{-2}$ during the 4.5-week period that transpired between Visit 2 and Visit 4, spiking to 3.3-4.3x$10^{15}$ $cm^{-2}$ only on Visit 5 thus, resulting in an $S_2$ gas density level 2-4 times that observed in the previous year. Due to the variation in the fitted $SO_2$ column density, the corresponding percent $S_2/SO_2$ ratios varied from relative low levels ~ 1-9 % on Visits 2 and 3, to substantially higher levels ranging from 2-30% to 14-43 % during Visits 4 and 5 when lower $SO_2$ column densities were detected.

*4.1.2 Dust scattering at Pele*

Quantitative characterization of the dust plumes seen in the ACS images is dependent on the careful assessment of the red leak and scattered light from Io's disk, and only qualitative analysis has been done so far. The density and/or distribution of dust emitted at the Pele vent appears to be variable, for example the brightness of the dust scattering detected above the Pele vent on Visit 4 is significantly less than that observed on Visit 5, consistent with the larger gas abundances seen on visit 5. Quantifying variations in the altitude of the Pele plume from the ACS images is also difficult because on some occasions the erupting plumes appear to overlap, i.e. given the line of sight angle of the observations, plumes whose longitudes may be different, but whose latitudes are fairly similar appear stacked one in front of the other within the 2-D images. For example, on Visit 1 when two different $S_2$ gas sources were detected in the STIS spectra, only one large scattering mass was observed in the ACS images in the region extending from just south of the Pele vent at 18 S to just north of the equator (Figs. 2 & 7). A similar issue arises on Visit 2. In this case, the scattering observed on the southern hemisphere is not evenly distributed about the Pele vent, instead the altitude of the plume extends higher north of the vent than it does on the south side of the vent. While it is possible that the Pele eruption was lopsided, an alternate reason for the observed spatial distribution would be a second plume source northward of the Pele plume (see discussion below).

On visits 3 thru 5 the Pele plume is significantly more isolated, on these visits the brightest regions within the plume, respectively, extend 7, 8, and 7 pixels along Io's limb and 5, 4, and 5 pixels above Io's limb (see Figure 2). Given that the plate scale of the ACS images is 0.027" /pixel corresponding to ~ 90 km/pixel on Io, for visits 3 thru 5 the brightest regions of the Pele plume were roughly 630-720 km wide and 360-450 km high. These values are consistent with that derived from the STIS data, where a height of ~320 ± 160 km is estimated for each plume. The large uncertainty in the estimate of altitude from the STIS data is because the region above the observed Pele plume was not imaged due to pointing errors, and so height estimates are based solely on observations at only 1 slit position; additionally, it is not clear if in the case of the weaker plume detections the plume filled the entire slit.

INSERT

FIG. 7

***4.2 Trends at Plumes other than Pele***

*4.2.1 Plumes detected on the 250 W limb*

Although the brightness of the plumes observed in the ACS images has not been quantified, for several of the regions in which significant dust scattering was recorded, spectra were also recorded in the STIS observations (see Fig. 4, 5 and 7). For example, on Visit 1 dust scattering was observed that extended northward of the Pele vent to just north of the equator. While it is likely that some portion of the observed signal results from particulates erupted at Pele, the plume appears to be centered at 6 S (see Fig. 7). Transmission spectra obtained from the corresponding Visit 1 STIS observations further indicate that $SO_2$ and $S_2$ gas was present in the slit covering above 0-10 S latitude and also centered near ~ 6 S. Within this pixel the longitude of the limb varied from 262-264 W at the time of the STIS observation, so any plume whose height was 200 km or more, would have been visible above the limb assuming that the source of the plume was within 25° of the limb. On Visit 2 dust scattering was also observed north of Pele extending to ~ 4 S; unfortunately, we did not obtain a transmission spectrum of this region on this visit. By Visit 3, the $SO_2$ absorption signature evident in the spectrum of the region centered near 6 S had diminished, and $S_2$ gas absorption and dust scattering signatures were only marginally observed. Because the spectrum noise at short wavelengths was high detection of the $SO_2$ gas during this observation is tentative. Fitting of the observed spectrum indicates that, the column density of $SO_2$ vented in that region dropped by at least a factor of 2 between Visits 1 and 3.

There was also evidence for dust and gas plume activity on the northern hemisphere, but limitations in coverage made correlation of the dust and gas signatures difficult. As Figure 7 shows, dust scattering was observed in the F250W filter in the region extending from 0 N to 30 N on Visit 2. Dust scattering that was less extensive both vertically and horizontally was also observed in this filter at locations overlapping this region on Visit 3 extending from 0 to 20 N (i.e. centered near 10 N), and on Visit 5 extending from 10 N to 30 N (i.e. centered near 18 N). However, evidence for corresponding gas absorptions at the location of the dust emission observed in the F250W filters on Visits 2 and 5 can not be corroborated because pointing errors prevented us from obtaining a transmission spectrum of the emitting regions (see Figure 4). On Visit 3 a spectrum of the region near 10 N, 260 W was obtained, in this case $SO_2$ gas was positively detected, while $S_2$ gas was only marginally detected. Given the wide extent of the scattering mass observed on the 250 W limb in ACS images on Visit 1 a dusty plume may have been active near 10 N;

however, on that visit no gas was positively detected in the spectrum taken near 10 N. Similarly, on Visit 4 dust scattering was marginally detected in the F250W filter in the region extending from 5 N to 20 N; on this visit $SO_2$ and $S_2$ gas absorptions were only tentatively detected in STIS data, and then only within the pixel extending from 11 to 28 N, and centered near 18 N (see Figures 4 and 5).

In general, regardless of what hemisphere a plume was detected on we find that wherever $S_2$ gas absorption was positively detected, significant dust scattering was also observed in the F250W filter. Similarly, with the exception of the 10 N plume observed on Visit 3, if $S_2$ gas absorption was not positively detected then significant dust scattering was not detected in the F250W filter. While there is no obvious active plume candidate for the 10 N and 6 S plume observations, gas detections obtained in the STIS pixel extending spatially from 11 N to 28 N may have resulted from activity at Daedalus Patera which extends from ~15 to 23 N, and 271 to 276 W and is centered near 19 N, 274 W. Detection of a gaseous plume in the STIS images from this location required a plume height extending > 10 km above the limb, but because Daedulus was 23° forward of the limb at time at which the F250W ACS observation the plume would have been observable in scattered light only if it extended > 163 km above Io's limb.

*4.2.2 Plumes detected on the 70 W limb*

Finally, on Visit 5 significant dust scattering was also detected on the limb opposite to the Pele plume near 70 W longitude. Unfortunately, STIS observations were taken only on the limb where the Pele plume could be imaged, so spectral signatures of the plumes on limb near 70 W longitude were not simultaneously acquired. Nevertheless, definitive plume activity was evident in the F250W filter on the 70 W limb on Visit 5 near the equator at 40 S, and 45 N. This activity is likely to have been associated with the eruption of Hi'iaka (2 S, 79 W), Masubi (45 S, 55 W) and Zal (37 N, 78 W). At the time of this observation both the Hi'iaka and Zal eruption zones were located on the front of observed Io hemisphere, while the Masubi vent would have been located a mere 15° degrees behind the 70 W limb requiring only that the plume extend to altitudes greater than 50 km to have been seen above the limb in scattered light. Scattering recorded in the F220W filter indicates that there may also have been activity northwest of the Zal paterae on Visit 3. Additionally, off-limb flux is evident near 45 N in the F250W filter images taken on Visits 1-3 and may be indicative of activity at Masubi, but the distinction of actual plume activity is harder to verify in these images. Detailed reduction is needed to verify the detection of plumes at these locations.

## 5.0 Discussion:

### *5.1 Significance of observed Pele behavior:*

The major goal of the HST Pele observing campaign completed in 2003 and 2004 was to determine whether the gas compositions derived from 1999 observation in which $S_2$ gas was first positively detected in Pele were typical. For the six dates of observations obtained between 1999 and 2004, the highest overall $S_2$ and $SO_2$ densities were observed in 1999 (Table III, Figure 6). $SO_2$ densities observed at Pele in the 2003 – 2004 observations were 2 to 5 times lower than that observed in 1999, and the $S_2$ densities were 5 to 10 times lower than the $S_2$ density first observed at Pele. Additionally, modest variations in the $SO_2$ and $S_2$ gas densities were observed at Pele in the 2003 and 2004 that were not well correlated; the highest $SO_2$ gas density was detected during Visit 3, while the highest $S_2$ gas density was observed during Visit 5. Although the $SO_2$ and $S_2$ gas densities observed at Pele in 1999 were substantially higher than that observed in 2003 and 2004, the $S_2/SO_2$ ratio inferred from the 1999 data corresponds to the median of the $S_2/SO_2$ ratios observed at Pele in 2003 and 2004—suggesting that at least the $S_2/SO_2$ ratio observed at Pele in 1999 was not atypical. Still, a more temporally extensive observing campaign would be needed to determine whether the $S_2/SO_2$ gas ratios observed in 2003 and 2004 represent the full range of $S_2/SO_2$ gas ratios produced near Pele.

Zolotov and Fegley (1998, 1999, 2000, 2001) have shown that the composition of the gas species within active plumes is dependent on temperature and the oxygen fugacity of the magma with which the gas is equilibrated in the vent. Therefore, eruption conditions can be evaluated from data on the composition of the eruption species, and it follows that changes in the composition of the eruption species are directly related to changes in the eruption conditions. Interestingly, based on measurements of the $SO_2/S/SO/S_2$ gas ratios derived from the 1996 HST/FOS and 1999 HST/STIS observations of Pele (McGrath *et al*. 2000, Spencer *et al*. 2000) Zolotov and Fegley first estimated vent temperatures of 1440±150 K, and an oxygen fugacity of the magma at Pele ranging from 2 to 3 log units below the nickel-nickel oxide buffer. Detailed analysis of Galileo and Cassini (Radebaugh *et al.* 2004) data suggests that Pele's magma temperature was consistently between 1300 and 1400 K over long time periods, with sporadic detections of higher (1400-1600 K) or lower (1200-1400 K) temperatures, thus confirming the magma temperatures first calculated by Zolotov and Fegley based on the 1999 and 1996 HST Pele observations.

Zolotov and Fegley also indicated that if the temperature at which the gases equilibrated is known then the range of the O/S ratios of the volcanic gases within the active plumes can be estimated to first order from the observed $S_2/SO_2$ ratios, as these are the major O and S bearing species. Assuming a magma temperature of 1440 K and the same oxygen fugacities derived for Pele's gas species based on the 1996 HST FOS data (McGrath et al. 2000, Zolotov and Fegley 2000, 2001) an O/S ratio of 1.1 or greater is implied by the 1999-2004 observations. If the 2003 and 2004 results are used in conjunction with the range of vent temperatures implied by the Galileo data, then the range of abundance ratios for the minor O and S bearing gas species within the Pele plume can also be predicted.

Similarly, the vent pressure of the 2003 and 2004 eruptions can also be estimated if the compositional measurements derived from the STIS data are used in conjunction with previously observationally derived magma temperatures and oxygen fugacities. Again, assuming the oxygen fugacities implied by 1996 observations, and a vent temperature of 1440 K then vent pressures of $1.0 \times 10^{-7}$-$5.0 \times 10^{-6}$ bar and $5.0 \times 10^{-6}$-$6.3 \times 10^{-5}$ bar are indicated for $S_2/SO_2$ ratios of 0.01-0.1 and 0.1-0.4 (or $SO_2/S_2$ of 100-10 and 10-2.5), respectively; or for a vent temperature of ~ 1970 K (the highest vent temperature tolerated by the analysis of the Galileo NIMS data (Lopes *et al*. 2001), the vent pressure ranges from $1.0 \times 10^{-1.0}$-$1.2 \times 10^{-1.0}$ bar and $1.2 \times 10^{-1.0}$-$7.9 \times 10^{-1.0}$ bar (Zolotov and Fegley 2000, 2001). If the magma temperature and oxygen fugacity were constant at Pele then significant variation in the eruption pressures are implied.

Although, constancy of the oxygen fugacity in the Pele plume has not been observationally corroborated, SO is the primary constituent by which changes in the oxygen fugacity can be tracked (Zolotov and Fegley 2001). Predictions of the SO abundance as a function of the O/S ratio and the magma temperature have also been calculated by Zolotov and Fegley (1998a, 2000). Comparison of these numbers indicates that unless the $SO_2/S_2$ ratio is less than one, i.e. unless $S_2$ becomes the dominant species in the erupting gas then the change in the SO abundance is less than an order of magnitude. For the range of $S_2/SO_2$ ratios observed at Pele between 1999 and 2004 the predicted SO mole fraction ranges from 0.05 to 0.08, and the evaluation of the oxidation state of the gas species at Pele from the combined 1996 and 1999 observations remains valid.

As discussed above, the gas ratios are dependent on the vent conditions (temperature, pressure) and oxygen fugacity of the magma and any variation in these factors could lead to the observed variation in the gas ratios. Based on the above discussion observations so far seem to indicate that variation in the oxygen fugacity, and the magma temperature may be limited. However, the chemical make up of the observed plumes is additionally affected by the amount of $SO_2$ and $S_2$ gas entrained within the magma on its ascent from depth, and this quantity can be variable (Leone and Wilson 2001, Cataldo *et al*. 2001a). Magma volatile interaction modeling by Cataldo *et al*. (2001b, 2002) also indicates that both the temperature of the magma and the pressure of the eruption change as a function of the density of gases entrained within the magma. Consequently, the reason for the observed variability in the $SO_2$ and $S_2$ gas densities and the relative $S_2/SO_2$ gas ratios needs to be investigated, and may be better understood based on detailed modeling of volatile and magma interaction during the ascent of the magma to the surface. Additionally, direct measurement of the vent temperature in conjunction with the gas densities would contribute to constraining the correlation between changes in the vent conditions and changes in the observed gas compositions.

Independent of what the source of observed variability is, the Pele plume still qualifies as an $S_2$ rich plume with an $S_2/SO_2$ ratio ~ 2-60x greater than the upper limit inferred for the Prometheus plume from the analysis of 2001 HST spectra (Jessup *et al*. 2004).

***5.2 Significance of other plume detections:***

On Februrary 24, 2003 a plume bearing $S_2$ gas and dust in the region near 6 ± 5 S, 260±15 W was detected. Prior to this detection, $S_2$ gas had only been directly detected in the plume of the Pele volcano (Spencer et al. 2000). Indirect evidence for $S_2$ emission from plumes other than Pele is supported by the 1996-2001 Galileo observations which revealed localized red deposits near the vents of multiple volcanos (McEwen *et al*. 1998b, Geissler 2003, Geissler *et al*. 2004a). For example, Galileo images obtained in April (Fig. 1) 1997 showed red streaks flanking a fresh lava flow located NE of the Pillan caldera at 9.5 S, 243 W, and superimposed on Pele's red ring (McEwen *et al*. 1998b). Plume eruptions observed by Galileo near Pillan in June 1997, were obtained just 1 week prior to the dusty plume detected by HST at Pillan; this eruption was later inferred to have emanated from a linear fissure system located NE of Pillan and associated with the hotspots and lava flows detected by the Galileo Solid State Imaging instrument in the region extending from 9.5 to 11.5 S and 242 to 243 W (Geissler *et al*. 2004a, Williams *et al*. 2001, McEwen *et al*. 1998b, Lopes *et al*. 1999). Since the limb of Io was at 262-268 W in the STIS spectra on Visit 1, so that the Pillan vent was located 130-170 km behind the limb, a gaseous plume eruption from this location would have to have extended > 170 km above the limb to have been visible in the STIS spectra at the time of the observations. Given that dust detections extending from 18 S to just north of the equator were recorded on Io's 250 W limb in the F250W filter; and that this scattering plume was observed to extended ~350-550 km above the surface, assuming that the source originated from Pillan's longitude, Pillan cannot be ruled out as the source of the second $S_2$ gas plume detected by HST (see Fig. 7).

Notably, the range of $SO_2$ and $S_2$ densities, and $S_2/SO_2$ ratio detected in the 6 S plume falls within the same range observed at Pele (see Fig. 6); thus, as was true for Pele, the $S_2/SO_2$ ratio of this plume is 2-6 times greater than the upper limit set at Prometheus, a plume produced by the remobilization of $SO_2$ frosts by lava (Jessup *et al*. 2004, Keiffer *et al*. 2000, Milazzo *et al*. 2001). If Pillan is the source of the plume, the detection of $S_2$ gas further supports that the detection of red deposits near Pillan throughout the Galileo era, resulted from fallout from gases released during Pillan eruption events occurring during that time.

Assuming that the marginal $S_2$ gas detection made near 18±5 N, 260±15W is associated with Daedulus Paterae (19 N, 274 W), the weakness of the $S_2$ signature is consistent with the low level of red coloration detected in this lava/crater complex (or around any other volcanic centers with locations consistent with being the source of this plume), at least during the 1996 – 2001 period of the Galileo observations.

Along the same vein, $S_2$ gas was not detected in the spectra of the region near 10±5 N, 260±15W. While a narrow region of red-diffuse surface particulates was evident in all of the Galileo observations obtained of the region extending from the tip of Pele's red ring at the equator to ~ 20 N (McEwen *et al*. 1998b, Geissler *et al*. 2004a, Geissler 2003), no hot spots or corresponding red streaks (Figure 1) were detected in this region during the Galileo or Cassini eras, which jointly covered a time period extending from 1995 to 2002 (Rathbun *et al*. 2004, Geissler *et al*. 2004a and b). Instead, a complex of lava flows was evident at 10 N, 240 W. If these flows were active in 2003 they may have generated a small plume of remobilized (volatized) surface materials dominated by $SO_2$ frosts, this would be consistent with the $SO_2$ signature observed near 10±5 N, 260±15W, and the lack of a positive $S_2$ detection in the region.

**6.0 Conclusions:**

In February 2003, March 2003 and January 2004 Pele plume transmission spectra were obtained during Jupiter transit with Hubble's Space Telescope Imaging Spectrograph (STIS), using the 0.1 arcsec long slit and the G230LB grating. Contemporaneous with the spectral data, UV and visible-wavelength images of the plume were obtained in reflected sunlight with the Advanced Camera for Surveys (ACS) prior to Jupiter transit. From these observations we report the following conclusions

1) For the dates of observation $SO_2$ and $S_2$ gas absorption was consistently detected at Pele; similarly the plume was consistently detected in reflected solar light.
2) The $SO_2$ and $S_2$ gas absorption signature directly over Pele varied significantly on 7-14 day time scales
3) The corresponding percent $S_2/SO_2$ ratio at Pele was also variable:
   $S_2/SO_2$ ratios in the range of 1-10% were observed on March 6 & 13 2003.
   An $S_2/SO_2$ ratio in the range of 14-43% was observed 9 months later in January 2004.
4) The 8-30% ratio derived from the 1999 $S_2$ discovery observations corresponded to the median of the $S_2/SO_2$ ratio observed at Pele in March 2003 and January 2004.
5) In spite of temporal variability at Pele, the plume still qualifies as an $S_2$ rich plume, Pele's $S_2/SO_2$ ratio ~ 2-60x > the upper limit set at Prometheus
6) Pele was not the only $S_2$ bearing plume on Io's trailing hemisphere at low latitudes—a dusty $S_2$ bearing plume was clearly detected near 6 S, the corresponding $S_2/SO_2$ ratio was similar to that observed at Pele on 3/13/2003, and 2-6x greater than the upper limit set to the 2001 Prometheus observation.
7) Plume activity was detected via dust scattering on the 250 W limb near 10 N, but simultaneous spectral observation of this plume when the dust plume was most distinctive was not obtained.

8) Plume activity was detected via dust scattering on the 70 W limb at 40 S, and 45 N these locations are consistent with the Masubi (45 S, 55 W) and Zal (37 N, 78 W) eruption sites.

**Acknowledgements:**

We thank Tony Roman, our support astronomer at STScI. Funding for this project was provided by STScI, NASA grants NAG5-13350 and NAG5-10497.

## TABLE I: STIS OBSERVATION TABLE

| VISIT # | DATE | OBSERVATION | TIME | EXPOSURE TIME | LATITUDE LONGITUDE[a] | $S_2$ GAS DETECTION | $SO_2$ GAS DETECTION | ACS DUST DETECTION (90 MINUTES EARLIER) |
|---|---|---|---|---|---|---|---|---|
| 1 | 2003-02-24 | o6n5b1010 | 02:11:36 | 535 | 6 S, 262 | yes | yes | F250W [c] |
| 1 | 2003-02-24 | o6n5b1020 | 02:21:54 | 535 | 6 S, 264 | yes | yes | F250W |
| 1 | 2003-02-24 | o6n5b1030 | 02:32:13 | 130 | 10 N, 265 | marginal | marginal | F250W |
| 1 | 2003-02-24 | o6n5b1040 | 02:35:25 | 130 | 10 N, 266 | marginal | marginal | F250W |
| 1 | 2003-02-24 | o6n5b1050 | 02:38:37 | 130 | 18 S, 266 | yes[b] | yes | F250W |
| 1 | 2003-02-24 | o6n5b1060 | 02:41:49 | 130 | 18 S, 267 | yes[b] | yes | F250W |
| 2 | 2003-03-06 | o6n5b2030 | 17:03:48 | 130 | 18 S, 264 | yes | yes | F250W |
| 2 | 2003-03-06 | o6n5b2040 | 17:07:00 | 130 | 18 S, 265 | yes | yes | F250W |
| 3 | 2003-03-13 | o6n5b3010 | 18:25:24 | 540 | 6 S, 260 | marginal | marginal | F250W [d] |
| 3 | 2003-03-13 | o6n5b3020 | 18:35:47 | 540 | 6 S, 262 | marginal | marginal | F250W |
| 3 | 2003-03-13 | o6n5b3030 | 18:46:11 | 130 | 10 N, 263 | marginal | yes | F250W |
| 3 | 2003-03-13 | o6n5b3040 | 18:49:23 | 130 | 10 N, 264 | marginal | yes | F250W |
| 3 | 2003-03-13 | o6n5b3050 | 18:52:35 | 130 | 18 S, 264 | yes | yes | F250 W |
| 3 | 2003-03-13 | o6n5b3060 | 18:55:47 | 130 | 18 S, 265 | yes | yes | F250 W |
| 4 | 2003-03-22 | o6n5b4010 | 15:26:30 | 540 | 3 N, 267 | no | no | none |
| 4 | 2003-03-22 | o6n5b4020 | 15:36:53 | 540 | 3 N, 269 | no | no | none |
| 4 | 2003-03-22 | o6n5b4030 | 15:47:17 | 130 | 18 N, 270 | marginal | marginal | F250W |
| 4 | 2003-03-22 | o6n5b4040 | 15:50:29 | 130 | 18 N, 270 | marginal | marginal | F250W |
| 4 | 2003-03-22 | o6n5b4050 | 15:53:41 | 130 | 18 S, 271 | yes | yes | F250W |
| 4 | 2003-03-22 | o6n5b4060 | 15:56:53 | 130 | 18 S, 271 | yes | yes | F250W |
| 5 | 2004-01-15 | o6n5d2010 | 19:16:01 | 495 | above 18 S, 262 | yes | yes | F250 W |
| 5 | 2004-01-15 | o6n5d2020 | 19:25:39 | 495 | above 18 S, 264 | yes | yes | F250 W |
| 5 | 2004-01-15 | o6n5d2050 | 19:41:42 | 130 | 18 S, 266 | ---------------------on Io surface----------------------- | | |
| 5 | 2004-01-15 | o6n5d2060 | 19:44:54 | 495 | 18 S, 266 | ---------------------on Io surface----------------------- | | |

a: latitude of the gas detection, longitude of the limb at the time of the STIS observations; the location of the limb in the ACS images recorded 90 minutes earlier is -15° W of the values listed for the STIS observations b: Although the absorption signature is weak, $S_2$ bands are clearly evident above 2700 Å; below 2700 Å detection of the $S_2$ bands is marginal c: A single scattering mass is observed on this date extending from 10 N to 20 S (see Figures 2 & 7).

d: scattering mass is observed at 6 S; however, the scattering signature overlaps that observed at the Pele vent on this date (see Figures 2 & 7).

## TABLE II: ACS OBSERVATION TABLE

UV OBSERVATIONS

| VISIT # | DATE | OBSERVATION[a] | TIME | EXPOSURE TIME (s) | FILTER | CENTRAL MERIDIAN LONGITUDE |
|---|---|---|---|---|---|---|
| 1 | 2003-02-24 | j6n5b1010 | 00:24:15 | 900 | F220W | 157 |
| 1 | 2003-02-24 | j6n5b1020 | 00:42:17 | 900 | F250W | 160 |
| 2 | 2003-03-06 | j6n5b2010 | 14:55:02 | 900 | F220W | 156 |
| 2 | 2003-03-06 | j6n5b2020 | 15:13:04 | 960 | F250W | 158 |
| 3 | 2003-03-13 | j6n5b3010 | 16:39:05 | 900 | F220W | 156 |
| 3 | 2003-03-13 | j6n5b3020 | 16:57:07 | 960 | F250W | 159 |
| 4 | 2003-03-22 | j6n5b4010[b] | 12:40:32 | 450 | F220W | 155 |
| 4 | 2003-03-22 | j6n5b4020 | 13:43:34 | 480 | F250W | 162 |
| 5 | 2004-01-15 | j6n512010 | 17:26:51 | 900 | F220W | 157 |
| 5 | 2004-01-15 | j6n512020 | 17:44:53 | 480 | F250W | 159 |

a: each of the ACS image pairs were geometrically corrected within the HST pipeline, and stored in a single image file.
b: Only one of the two exposures taken in this image pair was recorded uncorrupted, significantly reducing the signal-to-noise of this image (see Figures 2 & 7).

**TABLE III: SPECTRAL FITTING RESULTS**

| OBSERVATION | $SO_2$ (x $10^{16}$ $cm^{-2}$) | $S_2$ (x $10^{15}$ $cm^{-2}$) | $S_2/SO_2$ % | Assumed $SO_2$ temperature (K) |
|---|---|---|---|---|
| Pele 1999 $S_2$ Discovery | 4.0-10 | 8-12 | 8-30 | 300-500 |
| Pele – Visit 1 | 0.7-2.7 | 0.1-1.7 | 0.4 – 24 | 200-300 |
| Pele – Visit 2 | 2.3-4.5 | 0.7-2.0 | 1.6 - 8.7 | 200-300 |
| Pele – Visit 3 | 3.5-5.3 | 0.7-2.0 | 1.3 - 5.7 | 200-300 |
| Pele – Visit 4 | 0.7-2.7 | 0.5-2.0 | 1.8 – 29 | 200-300 |
| Pele – Visit 5 | 1.0-2.3 | 3.3-4.3 | 14 – 43 | 200-300 |
| 6 S – Visit 1 | 3.3-4.7 | 0.7-1.3 | 1.5 - 4.0 | 200-300 |
| 6 S – Visit 3 | 0.7-1.7 | < 0.7 | < 10 | 200-300 |
| 10 N – Visit 3 | 1.3-3.0 | < 0.7 | < 5 | 100-300 |
| 18 N – Visit 4 | 0.05-2.0 | < 1.3 | marginal | 100-300 |
| 2001 Prometheus-a[a] | 14.0-24 | < 1.0 | < 0.7 | 190-375 |
| 2001 Prometheus-b | 9.0-19 | < 1.0 | < 1.1 | 190-375 |

a: we include the 2001 Prometheus results for comparison purposes. Prometheus-a is the $SO_2$ column density range fit to the Prometheus data, Prometheus-b is the $SO_2$ column density range appropriate for the Prometheus observation independent of any contribution from the sublimation component of Io's atmosphere (see Jessup *et al*. 2004)

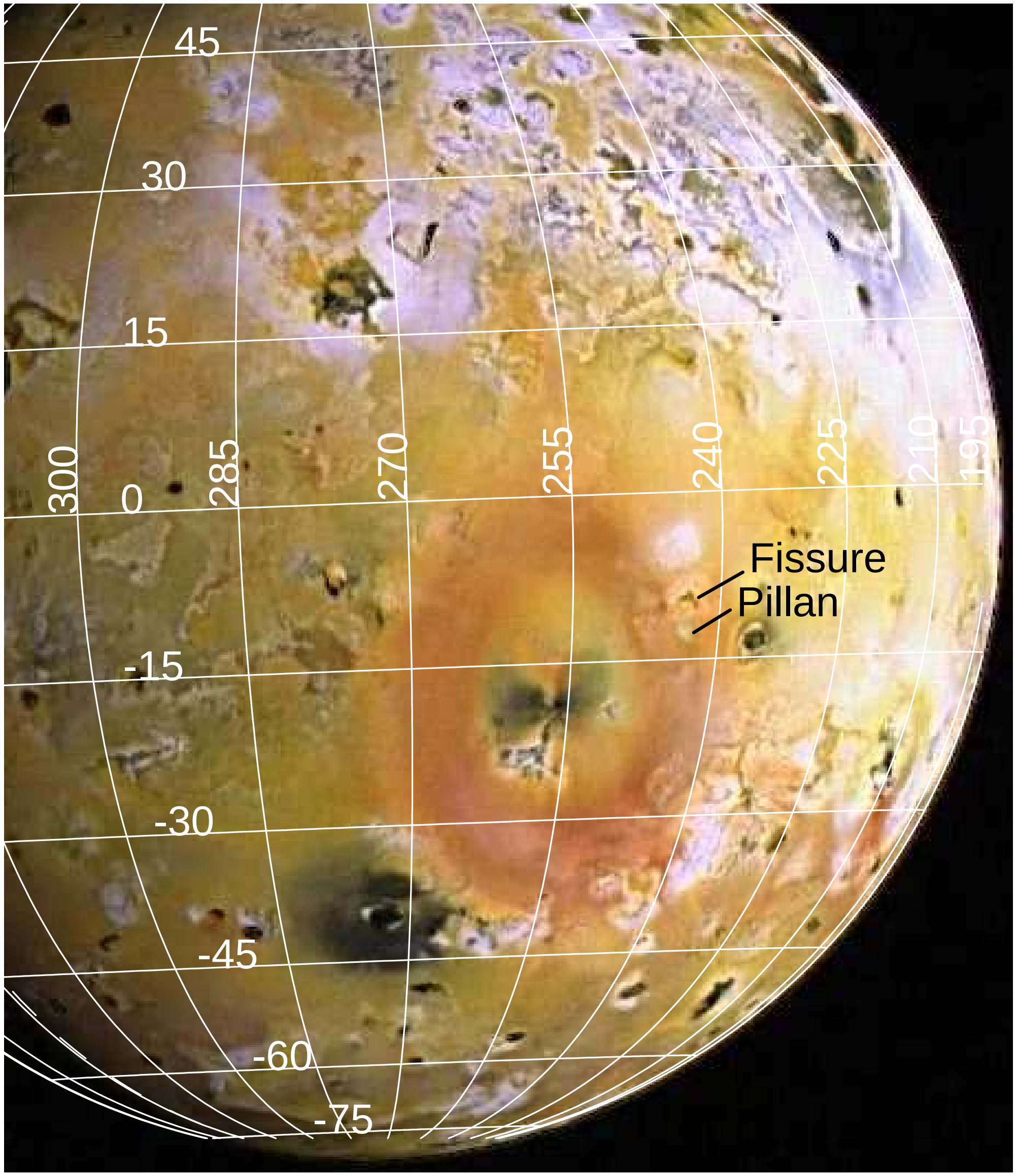

Figure 1: Galileo image of Pele's plume deposit region (Geissler 2003) taken in orbit 7 (April 03, 1997) just prior to the Pillan eruption. We highlight the location of the Pillan paterae and the fissure from which the 1997 Pillan eruption originated (Williams *et al.* 2001, Geissler *et al.* 2004a). Dust and gas were detected in our HST observations in the region extending from 20 S to 20 N, from longitudes of 249±20 W and 266 ±20 W, respectively—thus flanking Pele's fallout region.

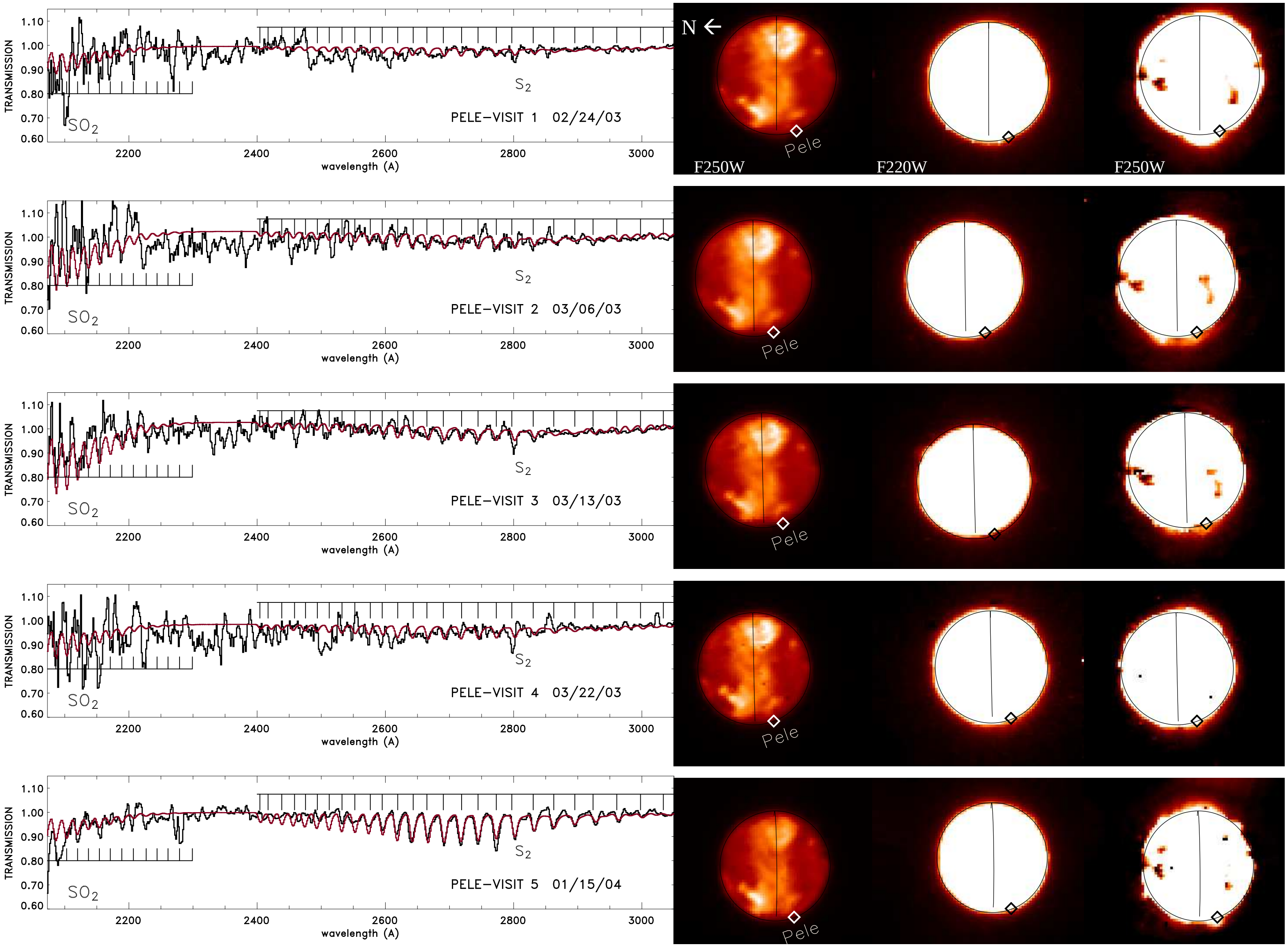

Figure 2: (left) HST/STIS transmission spectra (black) of the Pele plume, $SO_2$ and $S_2$ absorption bands are marked. We show the model transmission (red) fit to each of the observations assuming an $SO_2$ gas temperature of 250 K (see text for more details). We show on the right the HST/ACS images of Io obtained contemporaneously with the STIS data in the F220W (middle-images) and F250W (right and left -images). On the left we show the appearance of Io's disk (dominated by near-UV albedo patterns due to transmission of near-UV light longward of 2500 Å). In the middle and right panels we have subtracted a percentage of the mean radial profile from the image and applied a hard stretch to the images in order to emphasize the plumes above Io's limb. The location of the Pele vent is indicated by a diamond in each of the ACS images. It is clear that multiple plumes were active at the time of the observations.

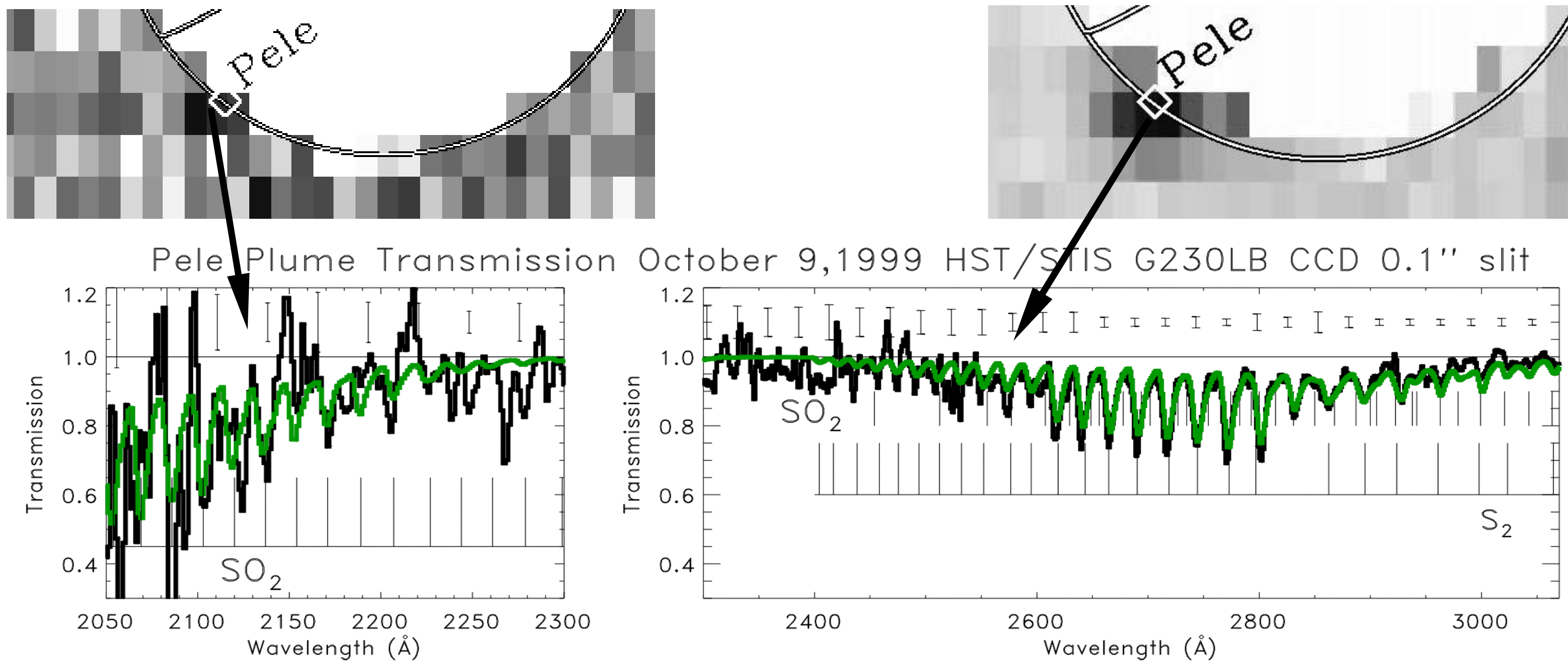

Figure 3: $S_2$ was first positively identified within the Pele plume in 1999 based on Hubble's Space Telescope Imaging Spectrograph (STIS) spectroscopy of Pele in Jupiter transit (Spencer *et al*. 2000). In the bottom panels we show the spectrum obtained above the Pele vent below 2300 Å (left) where the $SO_2$ gas absorption is strongest; and longward of 2400 Å (right) where both the strong $S_2$ and weak $SO_2$ gas absorption bands are present. Although in many instances absorption by the weak long wavelength $SO_2$ bands overlaps the strong $S_2$ absorption bands, absorption bands due solely to $SO_2$ are evident longward of 2850 Å. In the top panel we show gas absorption images produced by ratioing the mean reflectance at the bottom of the strong $SO_2$ (left) and $S_2$ (right) absorption bands to the mean reflectance in the continuum region between the bands. The location of Io's limb and the Pele vent are indicated.

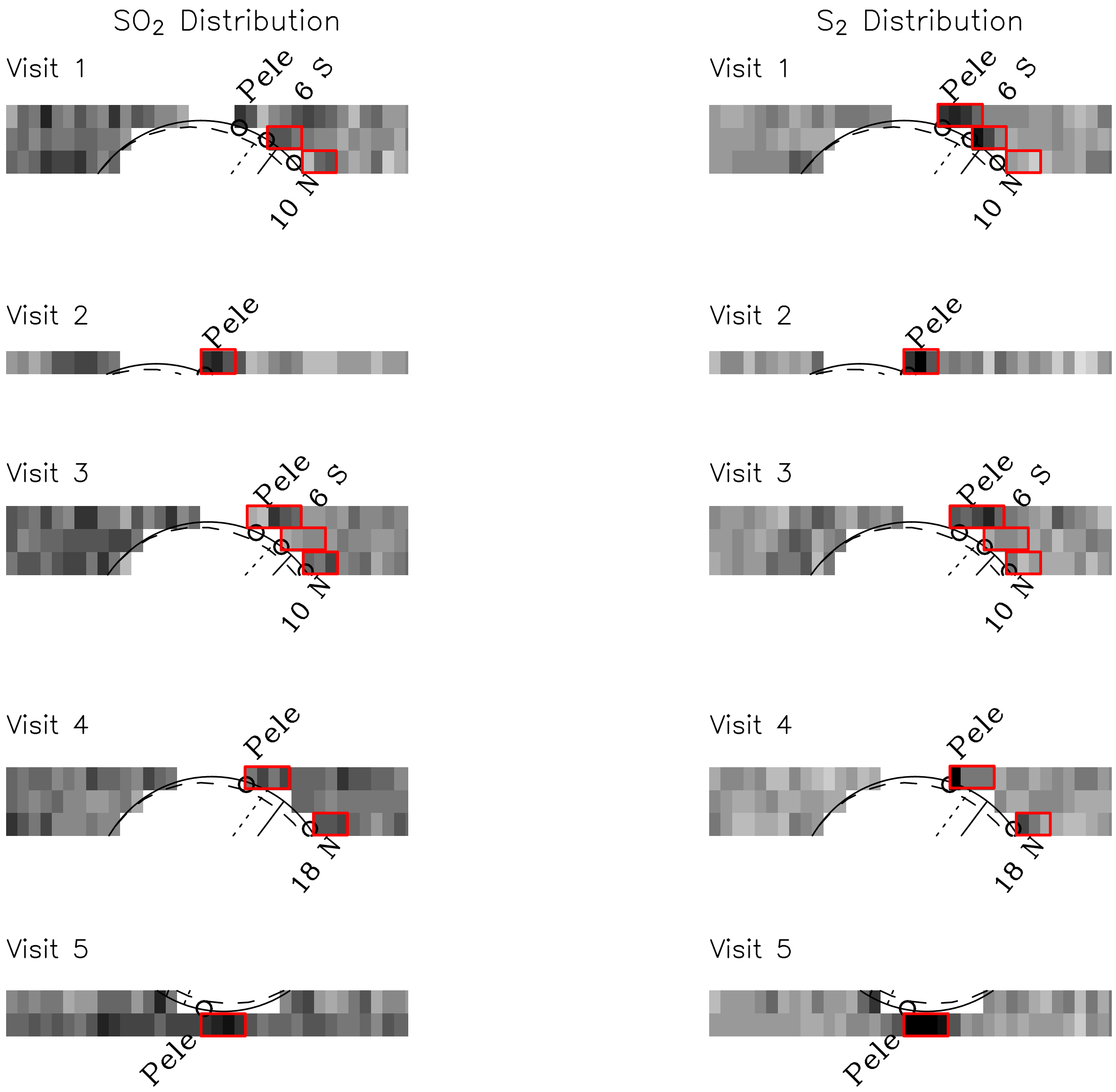

Figure 4: The spatial distribution of gas absorption derived from the STIS image cubes for the 2003 and 2004 observations, using the same methods described in Figure 3. The location of the Pele, 6 S, 10 N and 18 N gas detections are noted. At each of these locations a transmission spectrum was derived from the co-addition of the pixels outlined in red. The location of the Pillan lava flow is also noted by the intersection of the 243 W (dashed) meridian and the 9.5 S (dotted) latitude circle.

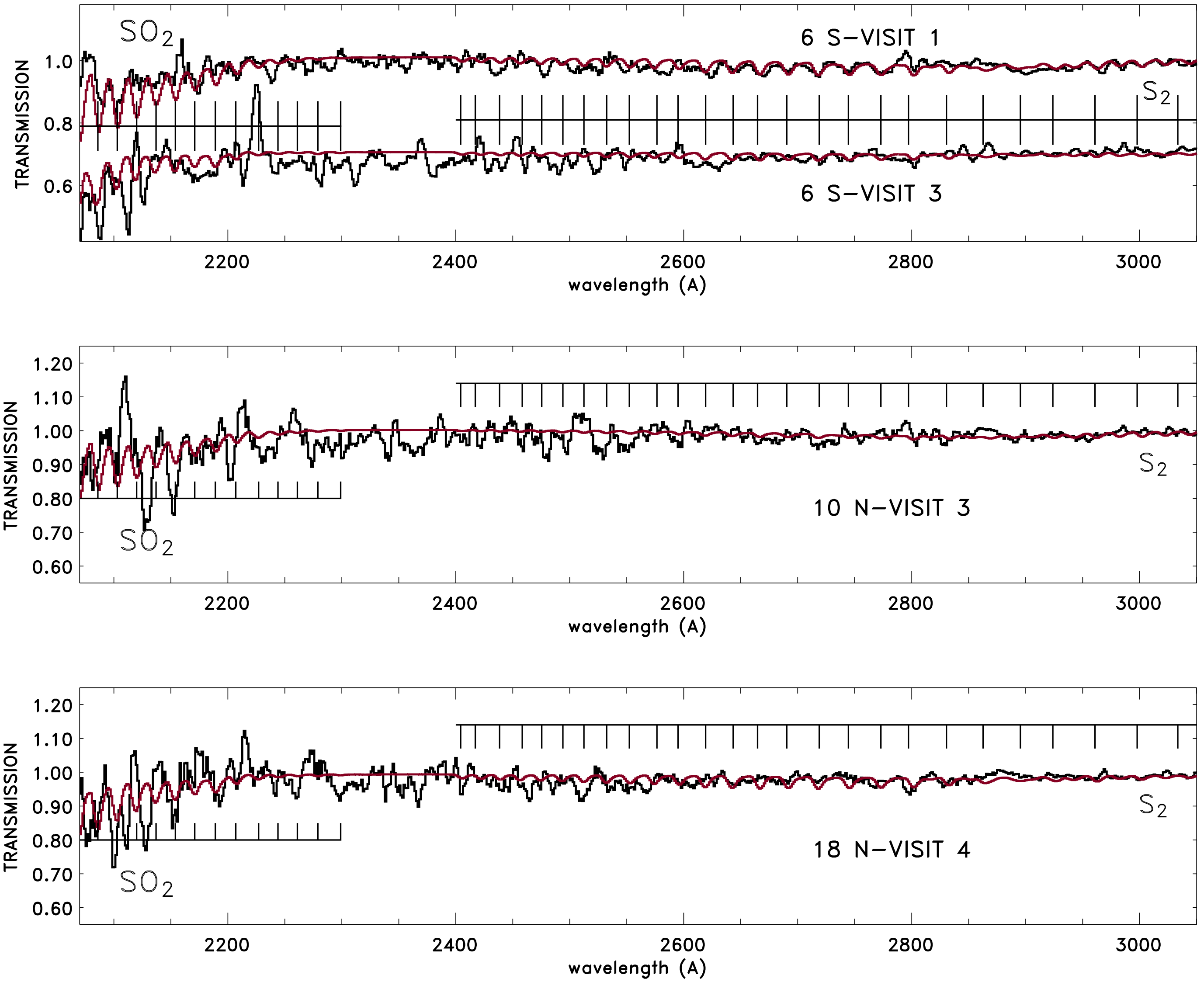

Figure 5: In addition to Pele, dusty plume detections were made near 6 S, 10 N, and 18 N (see Figures 2 and 7). We plot spectra showing the gas signature observed over each of these regions, and the model transmission spectrum (red) fit to each of the observations. $SO_2$ and $S_2$ gases were positively detected near 6 S during visit 1 on February 24, 2003; however, a spectrum taken of the same region during Visit 3 shows that the gas signatures had diminished. $S_2$ gas was not positively detected in the remaining spectra, and $SO_2$ gas was only positively detected in the spectrum obtained near 10 N on Visit 3.

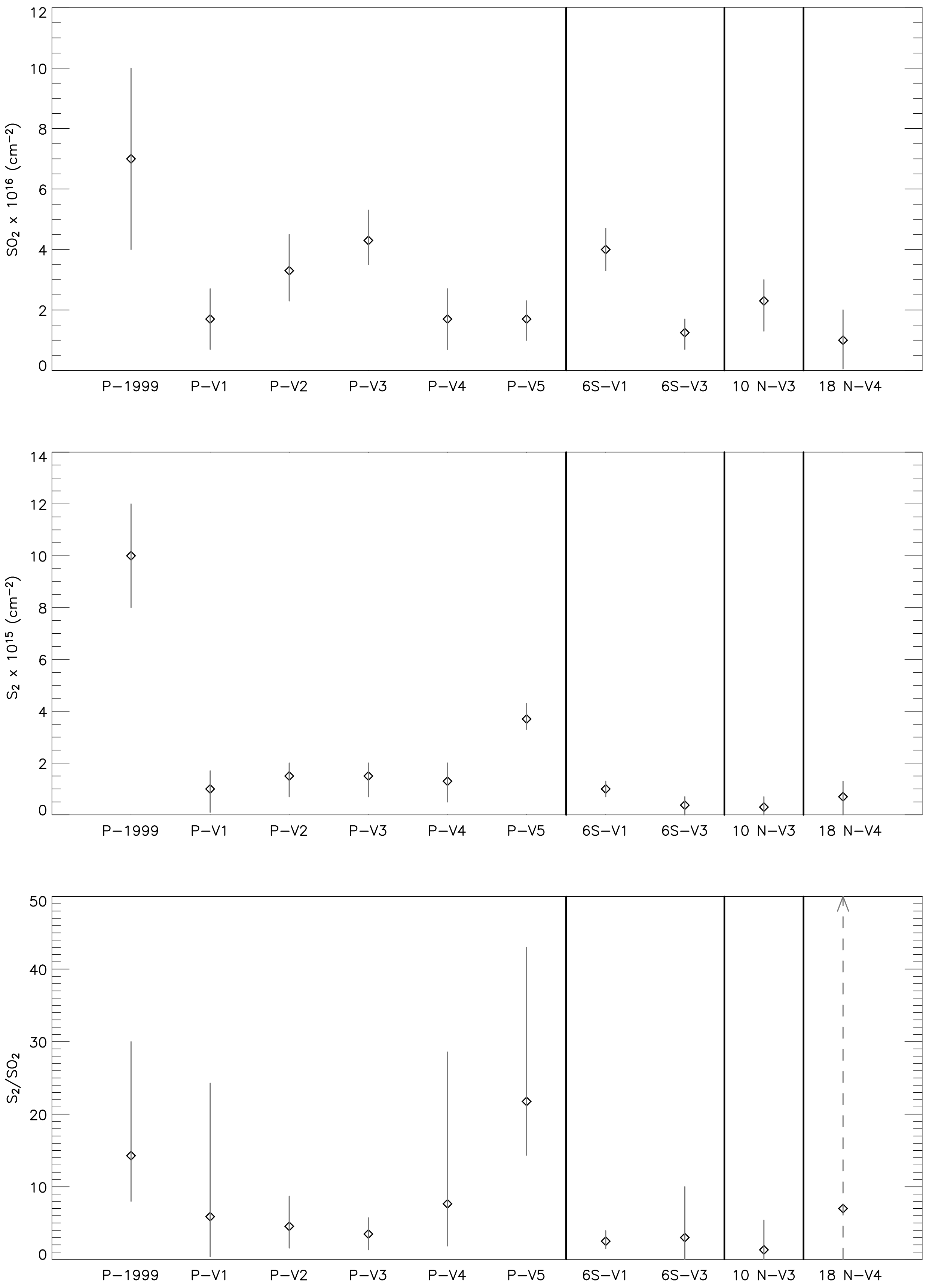

Figure 6: Plot of the $SO_2$ and $S_2$ densities inferred from the observed Pele (P) spectra obtained in 1999 and in visits 1-5 (V1-V5), 6S spectra obtained on V1 and V3, the 10 N spectra obtained on V3 and the 18 N spectrum obtained on V4.

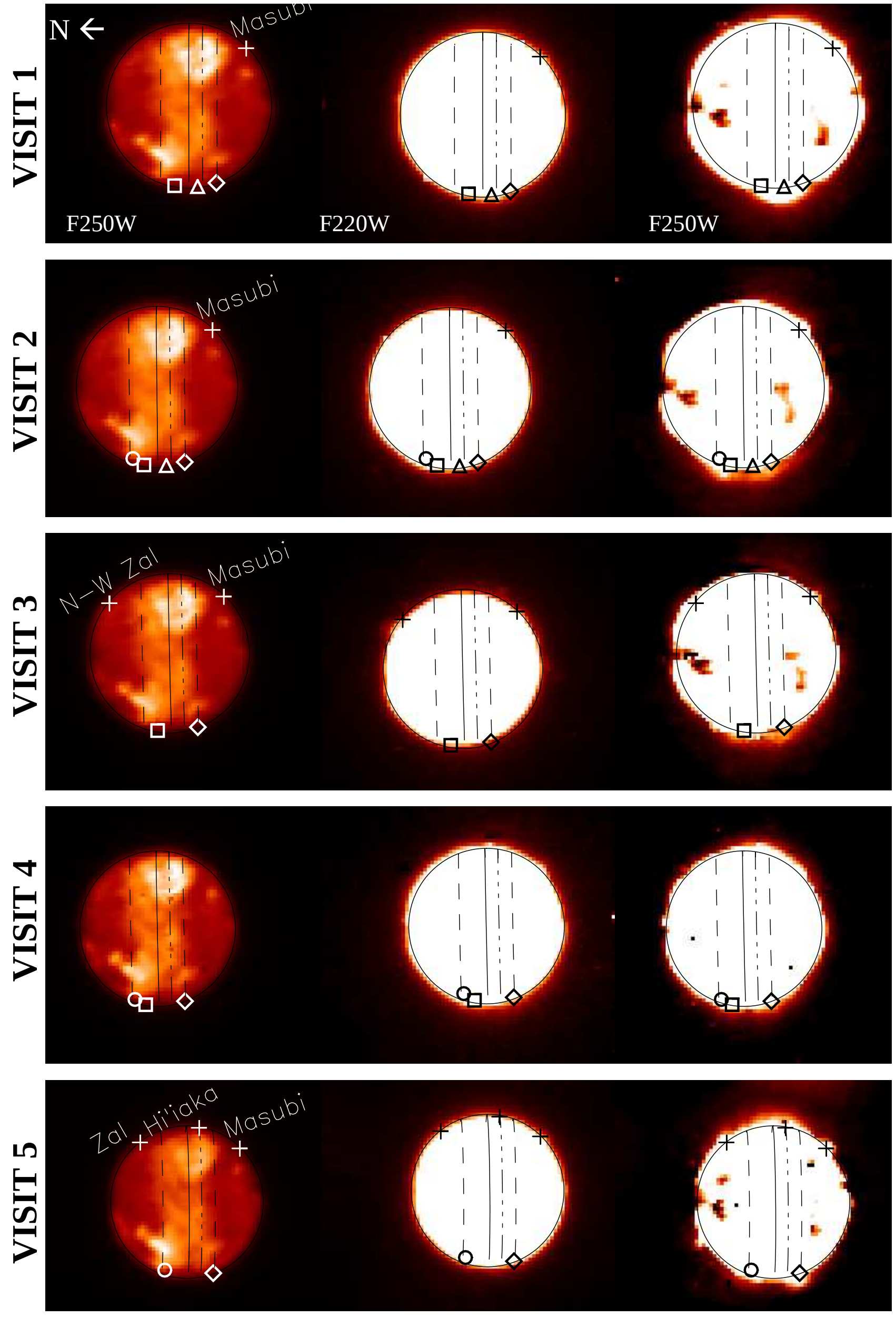

Figure 7: As in Figure 2, we show F220W (middle) and F250W (left and right panels) HST/ACS images of Io obtained contemporaneously with the STIS observations on visits 1 (top row) thru 5 (bottom row). Dust scattering is most prominent in the F250W filter. In this filter, a zone of high activity was evident between 20° N and S latitudes (dashed lines) on Io's limb near 250 W. In addition to Pele (diamond), dusty plumes were detected in the F250W filter near 6 S (triangle), 10 N (square) and 18 N (cirlce). These detections correspond to the gas detections observed along Io's limb at 260-270 W in the STIS data. The 10 N plume was prominent during Visit 2, and diminished during Visit 3. The 6 S plume was prominent during visit 1, diminished during Visit 2, marginal in Visit 3, and undetected in later visits. The 18 N plume may have been present during Visits 2 and 5, but is only marginally detected in the F250W filter on Visit 4. There are no known active thermal regions near 250 W± 15 at 10 N and 6 S latitude; however, the 18 N plume could be coincident with activity at Daedalus Patera (19 N, 274 W). Additionally, the 6 S plume also coincides with active lava flows observed in May thru September 1997 NE of the Pillan caldera (12 S, 243 W) at 9.5 S (dash-dot-dot-dot lines). Dust scattering was also recorded in the F250W filter on the 70 W limb on Visit 5 near the equator, and at 40° N and S latitude, this activity may have been associated with activity at Zal, Masubi and Hi'aka. Faint emissions recorded in the F220W filter near 45 N latitude, may be an indicator of activity northwest of the Zal paterae on Visit 3. Unfortunately, we did not obtain simultaneous STIS observations of the 70 W limb.